\documentclass[submission,copyright,creativecommons]{eptcs}
\providecommand{\event}{ACL2 2018} 
\usepackage{breakurl}             
\usepackage{underscore}           

\usepackage{booktabs}
\usepackage{proof}
\usepackage{epsfig}
\usepackage{float}
\usepackage{graphicx}
\usepackage{amsmath}
\usepackage{amsfonts}
\usepackage{amssymb}
\usepackage{xspace}

\newcommand{\smtlink}{\texttt{Smtlink}\xspace}
\newcommand{\Gtcp}{\ensuremath{G_{\mathrm{tcp}}}}
\newcommand{\Gsmt}{\ensuremath{G_{\mathrm{smt}}}}

\newcommand{\fsmt}{\ensuremath{\mathtt{f}_{\mathrm{smt}}}}
\newcommand{\facl}{\ensuremath{\mathtt{f}_{\mathrm{acl2}}}}
\newcommand{\hide}[1]{}

\newtheorem{theorem}{Theorem}[section]

\usepackage{listings}
\usepackage{color}

\usepackage{hyperref}
\hypersetup{
    colorlinks=true,
    linkcolor=blue,
    filecolor=magenta,      
    urlcolor=cyan,
  }
  
\definecolor{dkgreen}{rgb}{0,0.6,0}
\definecolor{gray}{rgb}{0.5,0.5,0.5}
\definecolor{mauve}{rgb}{0.58,0,0.82}

\definecolor{YanFavoriteColor}{rgb}{0.0,0.7,0.7}
\definecolor{MarkFavoriteColor}{rgb}{0.0,0.7,0.0}

\graphicspath{{./figs/}}

\lstdefinestyle{progstyle}{
    aboveskip=3mm,
    belowskip=3mm,
    showstringspaces=false,
    columns=flexible,
    basicstyle={\small\ttfamily},
    numbers=left,
    numbersep=5pt,
    numberstyle=\tiny\color{gray},
    keywordstyle=\color{blue},
    commentstyle=\color{dkgreen},
    stringstyle=\color{mauve},
    breaklines=true,
    breakatwhitespace=true,
    tabsize=3,
    keepspaces=true
  }

  \lstdefinestyle{snippet}{
    belowskip=0.5\baselineskip,
    basicstyle={\footnotesize\ttfamily},
    numbers=none,
    keywordstyle=\color{blue},
    commentstyle=\color{dkgreen},
    stringstyle=\color{mauve},
    showlines=true
}

\floatstyle{ruled}
\newfloat{Program}{htbp}{lop}[section]

\newlength{\codelinelen}
\newcommand{\codeline}[1]{%
  \setlength{\codelinelen}{\textwidth}%
  \addtolength{\codelinelen}{-4em}%
  \rule{2em}{0ex}\resizebox{\codelinelen}{!}{\texttt{#1}}}

\title{Smtlink 2.0}
\author{Yan Peng \qquad\qquad Mark R. Greenstreet
  \institute{
    University of British Columbia\thanks{This work is supported in part by the
      Institute for Computing, Information and Cognitive Systems (ICICS) at
      UBC.}\\
    Vancouver, Canada}
\email{\quad {yanpeng,mrg}@cs.ubc.ca}
}
\def\titlerunning{Smtlink 2.0}
\def\authorrunning{Yan Peng, Mark Greenstreet}

\begin{document}
\maketitle

\begin{abstract}
  \smtlink{} is an extension of ACL2 with Satisfiability Modulo Theories (SMT)
  solvers. We presented an earlier version at ACL2'2015. \smtlink{} 2.0 makes
  major improvements over the initial version with respect to soundness,
  extensibility, ease-of-use, and the range of types and associated
  theory-solvers supported. Most theorems that one would want to prove using an
  SMT solver must first be translated to use only the primitive operations
  supported by the SMT solver -- this translation includes function expansion
  and type inference. \smtlink{} 2.0 performs this translation using a sequence
  of steps performed by verified clause processors and computed hints. These
  steps are ensured to be sound. The final transliteration from ACL2 to Z3's
  Python interface requires a trusted clause processor. This is a great
  improvement in soundness and extensibility over the original \smtlink{}
  which was implemented as a single, monolithic, trusted clause processor.
  \smtlink{} 2.0 provides support for FTY \texttt{defprod}, \texttt{deflist},
  \texttt{defalist}, and \texttt{defoption} types by using Z3's arrays and
  user-defined data types. We have identified common usage patterns and
  simplified the configuration and hint information needed to use \smtlink{}.
\end{abstract}

\section{Introduction}

Interactive theorem proving and SMT solving are complementary verification techniques.
SMT solvers can automatically discharge proof goals with thousands to hundreds of
thousands of variables when the goals are within the theories of the solver.
These theories include linear and non-linear arithmetic, uninterpreted functions,
arrays, and bit-vector theories.
On the other hand, verification of realistic hardware and software systems
often involves models with a rich variety of data structures. Their proofs
involve induction, encapsulation (or other forms of higher-order reasoning),
and careful design of rules to avoid pushing the solver over an
exponential cliff of search complexity.
Ideally, we would like to use the capabilities of SMT solvers to avoid proving
large numbers of simple but tedious lemmas and combine those
results in an interactive theorem prover to enable the reasoning of properties
in large, realistic hardware and software designs.

\smtlink{} is a book developed for integrating SMT solvers into ACL2.
In the previous work~\cite{Peng2015-acl2}, we implemented an interface using a
large trusted clause-processor, and only supported a limited subset of SMT theories.
In this work, we introduce an architecture based on a collection of verified
clause processors that transform an ACL2 goal into a form amenable for discharging
with an SMT solver.  The final step transliterates the ACL2 goal into the syntax
of the SMT solver, invokes the solver, and discharges the goal or reports a
counter-example based on the results from the solver.  This final step is,
performed by a trusted-clause processor because we are trusting the SMT solver.
Because most of the translation is performed by verified clause processors, the
soundness for those steps is guaranteed.  Furthermore, the task of the final,
trusted clause processor is simple to describe, and that description
suggests the form for the corresponding soundness argument.
The modularized architecture is readily extensible which makes introducing new
clause transformation steps straightforward.

Our initial motivation for linking SMT solvers with ACL2 was for proving
linear and non-linear arithmetic for Analog and Mixed-Signal (AMS) designs~\cite{Peng2015-nfs}.
This original version supported the SMT solver's theories of boolean satisfiability
and integer and rational arithmetic, but provided no support for other types.
For example, to verify theorems involving lists, one would need to expand functions
involving lists to a user-specified depth, treat remaining calls as uninterpreted
functions which must return a boolean, integer, rational, or real.
The lack of support for a rich set of types restricted the applicability of
the original \smtlink{} to low-level lemmas that are primarily based on arithmetic.
The new \smtlink{} supports symbols, lists, and algebraic datatypes, with a
convenient interface to FTY types.

Other changes include better support for function expansion and uninterpreted functions.
The new \texttt{smtlink-hint} interface is simpler.
\smtlink{} generates auxiliary subgoals for ACL2 in ``obvious'' cases, such as
when the user adds a hypothesis for \smtlink{} that is not stated in the
original goal.
Hints for these subgoals share the same structure as the ACL2 hints and are
attached by the \smtlink{} hint syntax to the relevant subgoal -- no subgoal
specifiers are required!
When the SMT solver refutes a goal, the putative counterexample is returned to ACL2
for user examination.
Currently, \smtlink{} supports the Z3 SMT solver~\cite{Moura08} using Z3's Python API.
\smtlink{} is compatible with both Python 2 and Python 3, and can be used in ACL2(r)
with \texttt{realp}.

Documentation is online under the topic
\href{http://www.cs.utexas.edu/users/moore/acl2/manuals/current/manual/?topic=SMT____SMTLINK}{\texttt{:doc
    smtlink}}.
It describes how to install Z3, configure \smtlink{},
certify \smtlink{} and test the installation. It also describes the new
interface of the \smtlink{} hint, contains several tutorial examples, and some
developer documentation. Examples shown in this paper assumes proper
installation and setups are done as is described in \texttt{:doc
  smtlink}. It is our belief that given a more compelling argument of soundness
for the architecture, a richer set of supported types or theories, and a more
user-friendly user interface, \smtlink{} can support broader and larger proofs.

In Section~\ref{sec:arch} we present the architecture of \smtlink{} based on
verified clause processors, hint wrappers, and computed hints.
Section~\ref{sec:theories} describes the SMT theories supported by \smtlink{}
and sketches the soundness argument.
Section~\ref{sec:expl} gives a simple example where we use lists, alists, symbols, and booleans
to model the behavior of a ring oscillator circuit.
Section~\ref{sec:rlwk} provides a summary of related work, and we summarize the paper along
with describing ongoing and planned extensions to \smtlink{} in Section~\ref{sec:concl}.

\section{Smtlink 2.0 Architecture}\label{sec:arch}
This section describes the architecture of \smtlink{} 2.0.
We first introduce a running example to illustrate each of the steps taken by the clause processor.
Section~\ref{sec:arch.arch} describes the top-level architecture, followed by a brief description
of each of the clause processors used in \smtlink{}.

\subsection{An Nonlinear Inequality Example}\label{sec:arch.example}
To illustrate the architecture of \smtlink{},
we use a running example throughout this section.
Consider proving the following theorem in ACL2:

\begin{theorem}
$\forall x\in R$ and $\forall y \in R$, if $ \frac{9x^2}{8}+y^2 \le 1$ and
$ x^2-y^2 \le 1$, then $ y < 3(x-\frac{17}{8})^2 - 3$.
\end{theorem}

\begin{Program}[h]
  \caption{A nonlinear inequality problem}
  \label{prog:poly}
\begin{lstlisting}[language=Lisp]
(defun x^2-y^2 (x y) (- (* x x) (* y y)))

(defthm poly-ineq-example
  (implies (and (real/rationalp x) (real/rationalp y)
                (<= (+ (* (/ 9 8) x x) (* y y)) 1)
                (<=  (x^2-y^2 x y) 1))
           (< y (- (* 3 (- x (/ 17 8)) (- x (/ 17 8))) 3)))
  :hints(("Goal"
          :smtlink nil)))
\end{lstlisting}
\end{Program}

Program~\ref{prog:poly} shows the corresponding theorem definition in ACL2. To
use \smtlink{} to prove a theorem, a hint \texttt{:smtlink [smt-hint]} can be
provided using ACL2's standard \texttt{:hints} interface. \texttt{SMT::smt-hint}
is the hint provided to \smtlink{}. \texttt{SMT::smt-hint} follows a structure
that is described in
\href{http://www.cs.utexas.edu/users/moore/acl2/manuals/current/manual/?topic=SMT____SMT-HINT}{\texttt{:doc
    smt-hint}}. In this example, \texttt{:smtlink nil} suggests using \smtlink{}
without any additional hints provided to \smtlink{}.
The initial goal (underlining represented as clauses) is:

\begin{lstlisting}[style=snippet,language=LISP]
(IMPLIES (AND (RATIONALP X) (RATIONALP Y)
              (<= (+ (* (* 9 (/ 8)) X X) (* Y Y)) 1)
              (<= (X^2-Y^2 X Y) 1))
          (< Y (+ -3 (* 3 (+ X (- (* 17 (/ 8))))
                          (+ X (- (* 17 (/ 8))))))))
\end{lstlisting}

\subsection{The Architecture}\label{sec:arch.arch}

\begin{figure}[t]
  \begin{center}
   	\includegraphics[width=\textwidth]{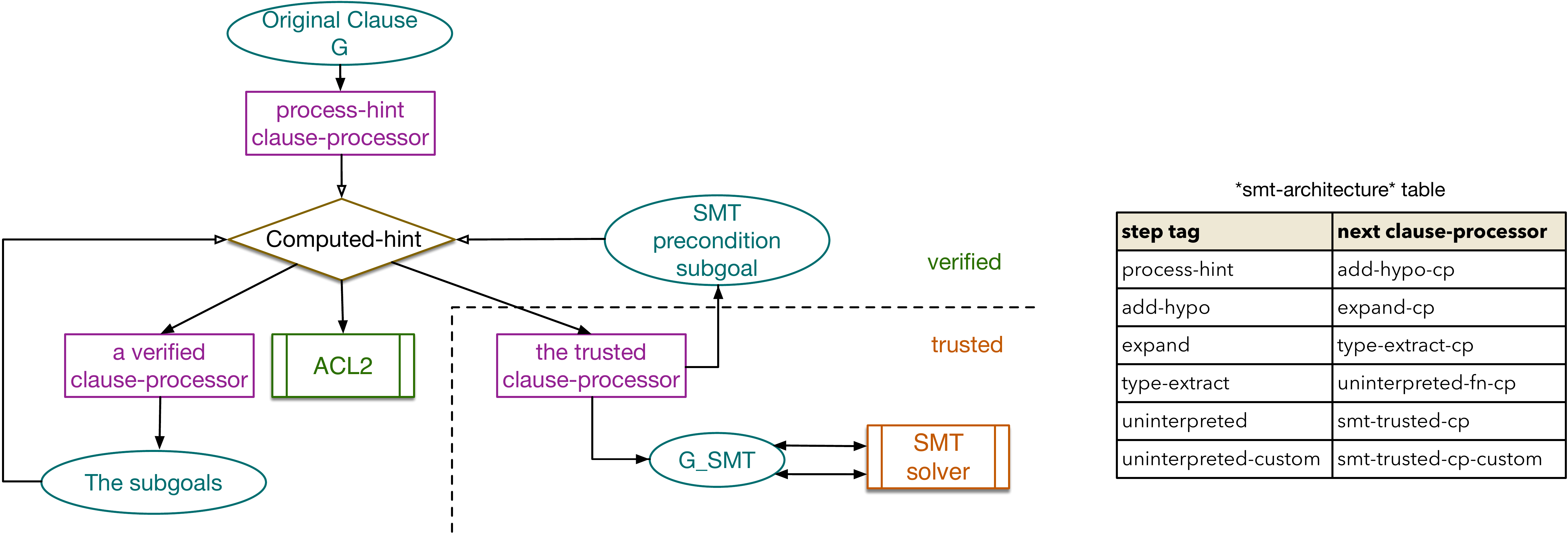}
    \caption[Smtlink Architecture]{\label{fig:arch} Smtlink Architecture}
  \end{center}
\end{figure}

Let $G$ denote the goal to be proven.
As shown in Figure~\ref{fig:arch}, the \smtlink{} hint invokes the verified
clause processor called \texttt{SMT::process-hint}.
The arguments to this clause-processor are the clause, $G$, and a list of hints provided by the user.
The \texttt{SMT::process-hint} performs syntactic checks on the user's hints and translates them into an
internal representation used by the subsequent clause processors.
The user can specify default hints to be used with all invocations of \smtlink{}.
These are merged with any hints that are specific for this theorem to produce \texttt{combined-hint}.
The \texttt{SMT::process-hint} adds a \texttt{SMT::hint-please} wrapper to the clause
and returns a term of the form\\
\codeline{`(((SMT::hint-please (:clause-processor (SMT::add-hypo-cp clause ,combined-hint))) ,@G))}

\smtlink{} 2.0 uses the hint wrapper approach from
\texttt{books/hints/hint-wrapper.lisp}.
In particular, we define a ``hint-wrapper'' called \texttt{SMT::hint-please} in package \texttt{smtlink}
that always returns \texttt{nil}.
Clauses in ACL2 are represented as lists of disjuncts, and our hint-wrapper
can be added as a disjunct to any clause without changing the validity of the clause.
\texttt{SMT::hint-please} takes one input argument -- the list of hints.

The computed-hint called \texttt{SMT::SMT-computed-hint} searches for a
disjunct of the form \\
\texttt{(SMT::hint-please ...)} in each goal generated by ACL2.
When it finds such an instance, the \texttt{SMT::SMT-computed-hint}
will return a \texttt{computed-hint-replacement} of the form:
\begin{lstlisting}[style=snippet]
  `(:computed-hint-replacement
    ((SMT::SMT-computed-hint clause))
    (:clause-processor (SMT::some-verified-cp clause ,combined-hint)))
\end{lstlisting}
This applies the next verified clause-processor called \texttt{SMT::some-verified-cp}
to the current subgoal and installs the computed-hint \texttt{SMT::SMT-computed-hint} on
subsequent subgoals again. The \texttt{SMT::some-verified-cp} clause-processor is one
step in a sequence of verified clause processors. \smtlink{} uses a
configuration table called \texttt{*smt-architecture*} to specify the sequence
of clause processors, as shown in Figure~\ref{fig:arch}.
Each clause processor consults this table to determine its successor.
By updating this table, \smtlink{} is easily reconfigured.

As described above, the initial clause processor for \smtlink{}, \texttt{SMT::process-hint},
adds a \\
\texttt{SMT::hint-please} disjunct to the clause to indicate that the clause processor,
\texttt{SMT::add-hypo-cp} should be the next step in translating the clause to a form
amenable for a SMT solver.

\hide{
\begin{lstlisting}[style=snippet,language=LISP]
(IMPLIES (NOT (SMT::HINT-PLEASE '(:CLAUSE-PROCESSOR (SMT::ADD-HYPO-CP CLAUSE ...))))
 (IMPLIES (AND (RATIONALP X) (RATIONALP Y)
               (<= (+ (* (* 9 (/ 8)) X X) (* Y Y)) 1)
               (<= (X^2-Y^2 X Y) 1))
          (< Y (+ -3 (* 3 (+ X (- (* 17 (/ 8))))
                          (+ X (- (* 17 (/ 8)))))))))
\end{lstlisting}

Notice how the next clause-processor is now embedded in the clauses wrapped with
\texttt{SMT::hint-please}. The clause-processor hint suggests that the next
clause processor to use is called \texttt{SMT::add-hypo-cp} and its second
argument stores the structured \texttt{SMT::smt-hint} to be used by \smtlink{} (which
is elided because of page limit). This leads us to the next section -- adding
user hypotheses.
}

\subsubsection{add-hypo-cp}
A key to using an SMT solver effectively is to tell it what it needs to know
but to avoid overwhelming it with facts that push it over an exponential complexity cliff.
The user can guide this process by adding hypotheses for the SMT solver -- typically
these are facts from the ACL2 logical world that don't need to be stated as
hypotheses of the theorem.
The clause-processor \texttt{SMT::add-hypo-cp} is a verified clause processor that adds
user-provided hypotheses to the goal.
Let $\mathbf{G}$ denote the original goal and $\mathbf{H_1}$, ... $\mathbf{H_n}$
denote the new hypotheses.
Each added hypothesis is returned by the clause processors as a subgoal to be
discharged by ACL2.
The user can attach hints to these hypotheses -- for example, showing that the
hypothesis is a particular instantiation of a previously proven theorem.
The soundness of \texttt{SMT::add-hypo-cp} is established by the theorem:
\begin{equation}
  \infer{G}{H_1\lor G & ... &  H_n \lor G & \bigwedge_{i=1}^N H_i \Rightarrow G}
\end{equation}
The term $\bigwedge_{i=1}^N H_i \Rightarrow G$ is the main
clause that gets passed onto the next clause-processor. In the following, let $G_{\mathrm{hyp}}$
denote $\bigwedge_{i=1}^N H_i \Rightarrow G$.

As for the example in Program~\ref{prog:poly}, the user didn't provide any
additional guidance. We see that the clauses generated are almost unchanged.
The only place that changed, is that \smtlink{} has installed the next
clause-processor to be \texttt{SMT::expand-cp}.

\begin{lstlisting}[style=snippet,language=LISP]
(IMPLIES (NOT (SMT::HINT-PLEASE '(:CLAUSE-PROCESSOR (SMT::EXPAND-CP CLAUSE ...))))
 (IMPLIES (AND (RATIONALP X) (RATIONALP Y)
               (<= (+ (* (* 9 (/ 8)) X X) (* Y Y)) 1)
               (<= (X^2-Y^2 X Y) 1))
          (< Y (+ -3 (* 3 (+ X (- (* 17 (/ 8))))
                          (+ X (- (* 17 (/ 8)))))))))
\end{lstlisting}

\subsubsection{expand-cp}
The clause-processor \texttt{SMT::expand-cp} expands function definitions to produce
a goal where all operations have SMT equivalents.  Function definitions are
accessed using \texttt{meta-extract-formula}. By default, all non-recursive
functions including disabled functions are expanded.
Recursive functions are expanded once and then treated as uninterpreted
functions. We attempt one level of expansion for recursive functions given the
reasoning that if the theorem can indeed be proved only knowing the return type
of the function, it should still be provable when expanded once.
If proving the theorem requires expansion of more than one level, or if all
occurrences of the function should uninterpreted, then we rely on the user to specify
this in a hint to \smtlink{}.
Let $\mathbf{G_{\mathrm{hyp}}}$ be the clause given to \texttt{SMT::expand-cp} and $\mathbf{G_{\mathrm{expand}}}$
be the expanded version.
The soundness of \texttt{SMT::expand-cp} is established by the theorem:
\begin{equation}
  \infer{G_{\mathrm{hyp}}}{G_{\mathrm{expand}} \Rightarrow G_{\mathrm{hyp}} & G_{\mathrm{expand}}}
\end{equation}
Presently, the clause $G_{\mathrm{expand}} \Rightarrow G_{\mathrm{hyp}}$ is returned to ACL2 for proof,
and $G_{\mathrm{expand}}$ is the main clause that gets passed onto the next clause-processor.

For the running example, function expansion produces the two subgoals. The main
clause that get passed onto the next clause-processor is:

\begin{lstlisting}[style=snippet,language=LISP]
(IMPLIES (NOT (SMT::HINT-PLEASE '(:CLAUSE-PROCESSOR (SMT::TYPE-EXTRACT-CP CLAUSE ...))))
 (IMPLIES (AND (RATIONALP X) (RATIONALP Y)
               (<= (+ (* (* 9 (/ 8)) X X) (* Y Y)) 1)
               (<= (LET NIL (+ (* X X) (- (* Y Y)))) 1))
          (< Y (+ -3 (* 3 (+ X (- (* 17 (/ 8))))
                          (+ X (- (* 17 (/ 8)))))))))
\end{lstlisting}

\noindent It tells \smtlink{} the next clause-processor is
\texttt{SMT::type-extract-cp} which does type declaration extraction. The other
subgoal is:

\begin{lstlisting}[style=snippet,language=LISP]
(IMPLIES
 (AND (NOT (SMT::HINT-PLEASE '(:IN-THEORY (ENABLE X^2-Y^2))))
      (IMPLIES (AND (RATIONALP X) (RATIONALP Y)
                    (<= (+ (* (* 9 (/ 8)) X X) (* Y Y)) 1)
                    (<= (LET NIL (+ (* X X) (- (* Y Y)))) 1))
               (< Y (+ -3 (* 3 (+ X (- (* 17 (/ 8))))
                               (+ X (- (* 17 (/ 8)))))))))
 (OR ... ;; some term essentially equal to nil
     (IMPLIES (AND (RATIONALP X) (RATIONALP Y)
                   (<= (+ (* (* 9 (/ 8)) X X) (* Y Y)) 1)
                   (<= (X^2-Y^2 X Y) 1))
           (< Y (+ -3 (* 3 (+ X (- (* 17 (/ 8))))
                           (+ X (- (* 17 (/ 8))))))))))
\end{lstlisting}

\noindent This second subgoal essentially proves that the expanded clause
implies the original clause. Notice how \texttt{SMT::hint-please} is used again
for passing other sorts of hints (that are not clause-processor hints) to ACL2
for help with the proof.

\subsubsection{type-extract-cp}
The logic of ACL2 is untyped; whereas SMT solvers such as Z3 use a many-sorted logic.
To bridge this gap, 
the clause-processor \texttt{SMT::type-extract-cp} is a verified clause processor for
extracting type information for free variables from the hypotheses of a clause.
\texttt{SMT::type-extract-cp} traverses the clause and identifies terms that syntactically
are hypotheses of the form \texttt{(type-p var)} where \texttt{type-p} is a known
type recognizer, and \texttt{var} is a symbol.
Let $G_{\mathrm{expand}}$ denote the clause given to \texttt{SMT::type-extract-cp};
$\mathbf{T_1}$, ... $\mathbf{T_m}$ denote the extracted type hypotheses;
$\mathbf{G_{\mathrm{expand}\backslash\mathrm{type}}}$ denote $G_{\mathrm{expand}}$ with the type-hypotheses removed; and
\begin{equation}
  G_{\mathrm{type}} = (\texttt{SMT::type-hyp}\ (\texttt{list}\ T_1 ... T_m) \texttt{:type}) \Rightarrow G_{\mathrm{expand}\backslash\mathrm{type}}
\end{equation}
where \texttt{SMT::type-hyp} logically computes the conjunction of the elements in the list
\texttt{(list $T_1$ \textrm{\ldots} $T_m$)} -- using \texttt{SMT::type-hyp} makes these
hypotheses easily identified by subsequent clause processors.
The soundness of \texttt{SMT::type-extract-cp} is established by the theorem:
\begin{equation}
  \infer{G_{\mathrm{expand}}}{G_{\mathrm{type}} \Rightarrow G_{\mathrm{expand}} & G_{\mathrm{type}}}
\end{equation}
$G_{\mathrm{type}} \Rightarrow G_{\mathrm{expand}}$ is the auxiliary clause returned back into ACL2 for
proof. $G_{\mathrm{type}}$ is the main clause that gets passed onto the next clause-processor.

For the running example, the main clause that gets passed onto the next
clause-processor is:
\begin{lstlisting}[style=snippet,language=LISP]
(IMPLIES (AND (NOT (SMT::HINT-PLEASE
                     '(:CLAUSE-PROCESSOR (SMT::UNINTERPRETED-FN-CP CLAUSE ...))))
              (SMT::TYPE-HYP (HIDE (LIST (RATIONALP X) (RATIONALP Y))) :TYPE))
 (IMPLIES (AND (<= (+ (* (* 9 (/ 8)) X X) (* Y Y)) 1)
               (<= (LET NIL (+ (* X X) (- (* Y Y)))) 1))
          (< Y (+ -3 (* 3 (+ X (- (* 17 (/ 8))))
                          (+ X (- (* 17 (/ 8)))))))))
\end{lstlisting}
\hide{
Now we have successfully isolated the type declarations from the main clause.
Two type declarations are extracted: \texttt{(RATIONALP X)} and
\texttt{(RATIONALP Y)}.}

\noindent The other auxiliary clause is:
\begin{lstlisting}[style=snippet,language=LISP]
(IMPLIES (AND (NOT (SMT::HINT-PLEASE
                     '(:IN-THEORY (ENABLE SMT::HINT-PLEASE SMT::TYPE-HYP)
                       :EXPAND ((:FREE (SMT::X) (HIDE SMT::X))))))
              (IMPLIES (SMT::TYPE-HYP (HIDE (LIST (RATIONALP X) (RATIONALP Y))) :TYPE)
                (IMPLIES (AND (<= (+ (* (* 9 (/ 8)) X X) (* Y Y)) 1)
                              (<= (LET NIL (+ (* X X) (- (* Y Y)))) 1))
                            (< Y (+ -3 (* 3 (+ X (- (* 17 (/ 8))))
                                            (+ X (- (* 17 (/ 8))))))))))
 (IMPLIES (AND (RATIONALP X) (RATIONALP Y)
               (<= (+ (* (* 9 (/ 8)) X X) (* Y Y)) 1)
               (<= (LET NIL (+ (* X X) (- (* Y Y)))) 1))
              (< Y (+ -3  (* 3 (+ X (- (* 17 (/ 8))))
                               (+ X (- (* 17 (/ 8)))))))))
\end{lstlisting}
This clause is returned for proof by ACL2.  The proof is straightforward -- essentially ACL2
simply needs to confirm that the terms we identified as type-hypotheses really are hypotheses
of the goal.

\subsubsection{uninterpreted-fn-cp}
Useful facts about recursive functions can be proven using the SMT solver's
support for uninterpreted functions.  As with variables, the return-types for
these functions must be specified.  For soundness, \smtlink{} must show that the
ACL2 function satisfies the user given type constraints.  This is done by
\texttt{SMT::uninterpreted-fn-cp}. Let $\mathbf{G_{\mathrm{type}}}$ denote the clause given to
\texttt{SMT::uninterpreted-fn-cp}, and let $\mathbf{R_1}$, ... $\mathbf{R_p}$ denote
the assertions about the types for each call to an uninterpreted function.
Let $\mathbf{Q_1}$ be the list of clauses
\begin{equation}\begin{array}{rcl}
  Q_1 &=& (\texttt{SMT::type-hyp}\ (\texttt{list}\ R_1)\ \texttt{:return}) \lor G_{\mathrm{type}}\ ,\ ...\ ,\\
      & & (\texttt{SMT::type-hyp}\ (\texttt{list}\ R_p)\ \texttt{:return}) \lor G_{\mathrm{type}}
\end{array}\end{equation}
Finally, let $\mathbf{G_{tcp}}$ be the clause
\begin{equation}
  G_{tcp} = \bigwedge_{i=1}^p (\texttt{SMT::type-hyp}\ (\texttt{list}\ R_i)\ \texttt{:return}) \Rightarrow G_{\mathrm{type}}
\end{equation}
The soundness of \texttt{expand-cp} is established by the theorem:
\begin{equation}
  \infer{G_{\mathrm{type}}}{Q_1 & G_{tcp}}
\end{equation}
The list of clauses $Q_1$ is returned to ACL2 for proof, and $G_{tcp}$ is tagged
with a hint to be checked by the trusted clause processor.

Our running example does not make use of uninterpreted functions in the SMT solver;
so, the clause $G_{tcp}$ is the same as $G_{\mathrm{type}}$ except for the detail
that the \texttt{SMT::hint-please} term now specifies that the next clause-processor
to use is the final trusted clause-processor \texttt{SMT::smt-trusted-cp}:
\begin{lstlisting}[style=snippet,language=LISP]
(IMPLIES (NOT (SMT::HINT-PLEASE '(:CLAUSE-PROCESSOR (SMT::SMT-TRUSTED-CP CLAUSE ... STATE))))
 (OR (NOT (SMT::TYPE-HYP (HIDE (LIST (RATIONALP X) (RATIONALP Y))) :TYPE))
     (IMPLIES (AND (<= (+ (* (* 9 (/ 8)) X X) (* Y Y)) 1)
                   (<= (LET NIL (+ (* X X) (- (* Y Y)))) 1))
              (< Y (+ -3 (* 3 (+ X (- (* 17 (/ 8))))
                              (+ X (- (* 17 (/ 8))))))))))
\end{lstlisting}

\subsubsection{smtlink-trusted-cp}

\begin{figure}[h]
  \begin{center}
   	\includegraphics[width=0.65\textwidth]{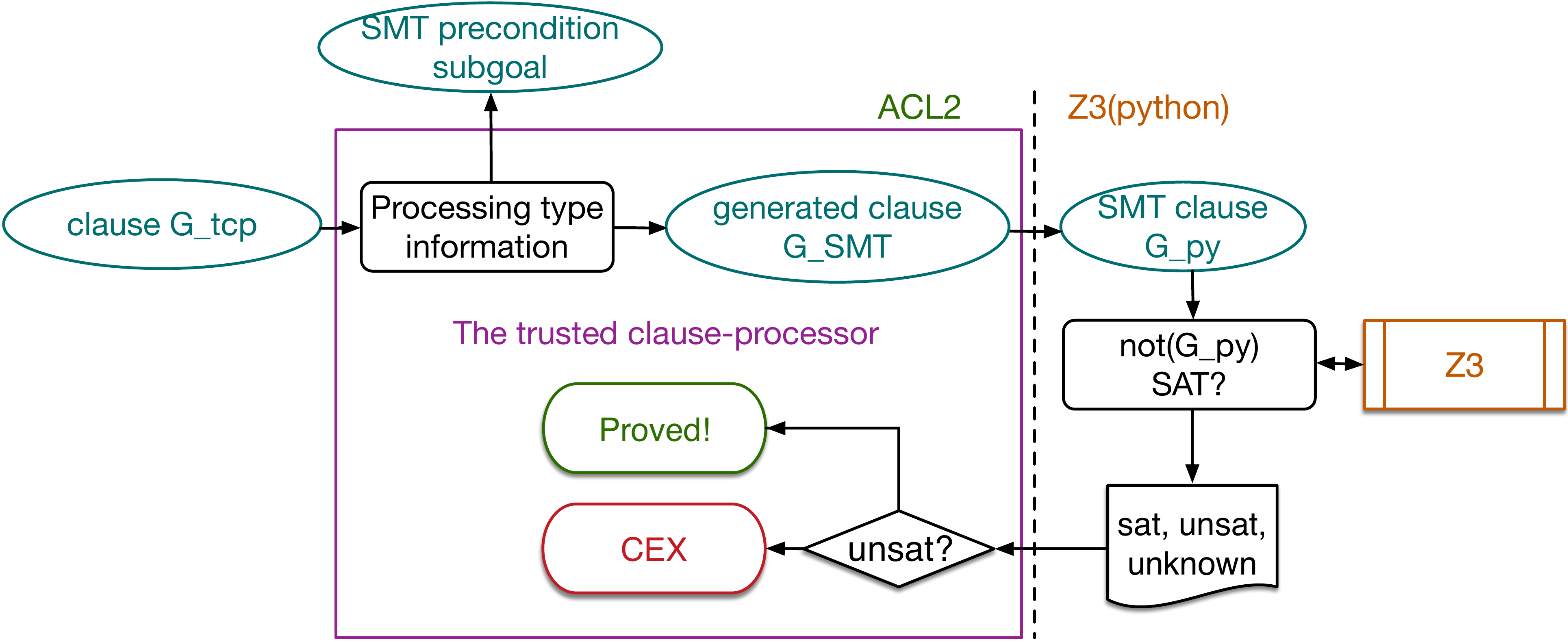}
    \caption[The trusted clause-processor]{\label{fig:trusted} The trusted clause processor}
  \end{center}
\end{figure}

Figure~\ref{fig:trusted} shows the internal architecture of the trusted
clause-processor, for which its correctness hasn't been proven.
\smtlink{} returns counter-examples generated by the SMT solver back into
ACL2.  The Python interface to Z3 for \smtlink{} includes code to translate a
counterexample from Z3 into list-syntax for ACL2. The form is not necessarily in
ACL2 syntax. We wrote a printing function that takes the Z3 counter-examples and
prints it out in an ACL2-readable form. These are all done through the trusted
clause processor. Currently, the forms returned into ACL2 are not evaluable. For
example, counter-examples for real numbers can take the form of an algebraic
number, e.g.
\begin{lstlisting}[style=snippet,language=LISP]
((Y (CEX-ROOT-OBJ Y STATE (+ (^ X 2) (- 2)) 1)) (X -2))
\end{lstlisting}
We plan to generate evaluable forms in the future. Learn more about
counter-example generation, see 
\href{http://www.cs.utexas.edu/users/moore/acl2/manuals/current/manual/?topic=SMT____TUTORIAL}{\texttt{:doc
    tutorial}}.

The trusted clause-processor returns a subgoal called ``SMT precondition'' back
into ACL2. By ensuring that certain preconditions are satisfied, we are able to
bridge the logical meaning for tricky cases, and therefore ensure soundness. We
will see explanations on this issue in Section~\ref{sec:theories}. For this
running example, there are no preconditions to be satisfied and the following
clause trivially holds:
\begin{lstlisting}[style=snippet,language=LISP]
(IMPLIES (AND (NOT (SMT::HINT-PLEASE
                     '(:IN-THEORY (ENABLE SMT::MAGIC-FIX SMT::HINT-PLEASE SMT::TYPE-HYP)
                       :EXPAND ((:FREE (SMT::X) (HIDE SMT::X))))))
              (NOT (AND (OR (NOT (<= (+ (* (* 9 (/ 8)) X X) (* Y Y)) 1))
                            (LET NIL T))
                        T)))
 (OR (NOT (SMT::TYPE-HYP (HIDE (LIST (RATIONALP X) (RATIONALP Y))) :TYPE))
     (IMPLIES (AND (<= (+ (* (* 9 (/ 8)) X X) (* Y Y)) 1)
                   (<= (LET NIL (+ (* X X) (- (* Y Y)))) 1))
              (< Y (+ -3 (* 3 (+ X (- (* 17 (/ 8))))
                              (+ X (- (* 17 (/ 8))))))))))
\end{lstlisting}

\subsection{A Few Notes about the Architecture}
The new architecture clearly separates what's verified from what's trusted.
As is shown in Figure~\ref{fig:arch}, given an original goal, the \smtlink{}
workflow goes through a series of verified clause-processor, which won't change
the logical meaning of the original goal. A computed-hint is installed to
provide hints for the next step, but won't change the logical meaning of the
goals either. The final trusted clause-processor takes a goal that logically
implies the original goal and derives information needed for the translation
through three sources of information it uses. First, information encoded in
the clause using \texttt{SMT::type-hyp}. This is sound because the definition of
\texttt{SMT::type-hyp} is a conjunction of the input boolean list. Second,
information about FTY types from \texttt{fty::flextypes-table}.
Presently, we trust this table to be correct and intact.
We plan to use rules about FTY types to derive this information in the future.
Third, we use information stored in \texttt{SMT::smt-hint} about
configurations, including what is the Python command, whether to treat integers
as reals, where is the \texttt{ACL2_to_Z3} class file and so on. We believe
this information can not accidentally introduce soundness issues from the user.
For experimenters and developers, we allow this low-level interface to be
overridden by the user using \texttt{:smtlink-custom} hint. Using such a
``customized'' \smtlink{} requires a different trust-tag than the standard
version, thus providing a firewall between experiments and production versions.
See \texttt{:doc tutorial} for how to use customizable \smtlink{}.



\section{Types, Theories, and Soundness}\label{sec:theories}
\smtlink{} translates the original goal, $G$ to an expanded goal, $\Gtcp$
for the trusted clause processor through a series of verified clause processors.
Thus, we regard the translation to $\Gtcp$ as sound.
The trusted clause processor translates $\Gtcp$ into a form that can be checked
by the SMT solver; we refer to this translated form as $\Gsmt$.
Let $x_1$, $x_2$, \ldots, $x_n$ denote the free variables in $\Gtcp$,
let $\Gsmt$ denote the translated goal,
and $\tilde{x}_1$, $\tilde{x}_2$, \ldots, $\tilde{x}_n$ denote
the free variables of $\Gsmt$.
For soundness, we want
\begin{equation}\label{eq:soundness}\begin{array}{rcl}
  \mathrm{SMT} \vdash \Gsmt &\Rightarrow& \mathrm{ACL2} \vdash \Gtcp
\end{array}\end{equation}
In the remainder, we assume
\begin{itemize}
  \item ACL2 and the SMT solver are both sound for their respective theories.
  \item The SMT solver is a decision procedure for a decidable fragment of first-order logic.
    In particular, this holds for Z3, the only SMT solver that is currently
    supported by \smtlink{}. In addition, we are working with a quantifier-free
    fragment of Z3's logic.
  \item There is a one-to-one correspondence between the free variables of $\Gtcp$
    and the free variables of $\Gsmt$.
    This is the case with the current implementation of \smtlink{}.
\end{itemize}

Now, suppose that $\Gtcp$ is not a theorem in ACL2.  Then, by G\"{o}del's
Completeness Theorem, there exists a model of the ACL2 axioms that satisfies
$\neg \Gtcp$. We need to show that in this case there exists a model of the SMT
solver's axioms that satisfies $\neg \Gsmt$.  There are two issues that we must
address. First, we need to provide, for the interpretation of any function
symbol $\facl$ in $\Gtcp$, an interpretation for the corresponding function
symbol $\fsmt$ in $\Gsmt$.
This brings us to the second issue: the logic of ACL2 is untyped, but the logic
of SMT solvers including Z3 is many-sorted.
Thus, there are models of the ACL2 axioms that have no correspondence with the
models of the SMT solver.
We restrict our attention to goals, $\Gtcp$ where the type of each subterm in the
formula can be deduced.  We refer to such terms as translatable.  If $\Gtcp$ is
not translatable, then \smtlink{} will fail to prove it.

For the remainder, we restrict our attention to translatable goals.
Because $\Gtcp$ is translatable, there is a set $R$ of unary recognizer functions
(primitives such as rationalp that return a boolean) and also a set $S$ of other
functions, such that every function symbol in $\Gtcp$ is a member of $R$ or of $S$,
and every function in $S$ is ``well-typed'' with respect to $R$ in some sense that
we can define roughly as follows.
We associate each function symbol $\facl$ in $S$ with a function symbol $\fsmt$
of Z3, and each predicate $r$ in $R$ with a type in Z3.
The trusted clause processor checks that there is a ``type-hypothesis'' associated
with every free variable of $\Gtcp$ and a fixing function for every type-ambiguous
constant (e.g.\ \texttt{nil}) -- $\Gtcp$ holds trivially if any of these
type-hypotheses are violated.
For every function $\facl$ in $S$, we associate a member of $R$ to each of its
arguments (i.e.\ a ``type'') and also to the result.
For user-defined functions (i.e.\ uninterpreted function for the SMT solver),
\smtlink{} generates a subgoal for each call to $\fsmt$: if the arguments satisfy
their declared types (i.e., predicates from $R$), then the result must satisfy its
declared type as well.
For built-in ACL2 functions (e.g.\ \texttt{binary-plus}) we assume the ``obvious''
theorems are present in the ACL2 logical world.
Now suppose we have a model, $M_1$, of $\neg \Gtcp$, and consider the submodel,
$M_2$, containing just those objects $m$ such that $m$ satisfies at least one
predicate in $R$ that occurs in $\Gtcp$. Note that $M_2$ is closed
under (the interpretation of) every operation in $S$, because $\neg \Gtcp$
implies that all of the ``type-hypotheses'' of $\Gtcp$ are true in $M_1$. This
essentially excludes ``bad atoms'', as defined by the function
\texttt{acl2::bad-atom}.
Then because $\Gtcp$ is quantifier-free, $M_2$ also satisfies $\neg \Gtcp$.
We can turn $M_2$ into a model $M_2'$ for the language of Z3, by assigning the
appropriate type to every object.  (As noted in Section~\ref{sec:reals}, $M_2'$
satisfies the theory of Z3 if $M_2$ is a model of ACL2(r); but for ACL2 that is
not the case, so in future work, we expect to construct an extension of $M_2'$
that satisfies all of the axioms for real closed fields.)
Then we have the claim: for every assignment $s$ from the free
variables of $\Gtcp$ to $M_2$ with corresponding typed assignment $s'$ from the
free variables of $\Gsmt$ to $M_2'$ , if $\neg \Gtcp$ is true in $M_2$
under $s$, then $\neg \Gsmt$ is true in $M_2'$ under $s'$.
Thus, if $\Gtcp$ is translatable, and $\neg \Gsmt$ is unsatisfiable, we conclude
that $\Gtcp$ is a theorem in ACL2.

In the rest of this section, we discuss for each of the recognizer functions and
each of the basic functions in ACL2, how we associate them with the
corresponding Z3 functions.

\subsection{Booleans, Integers, Rationals, and Reals}\label{sec:reals}
If a term is a boolean constant, then the translation to the SMT solver is direct.
Likewise, if $x_i$ is free in $\Gtcp$ and \texttt{(booleanp $x_i$)}
is one of the hypotheses of $\Gtcp$, then $\Gtcp$ holds trivially in the case that
$x_i \not\in \{\texttt{t},\,\texttt{nil}\}$.  Thus, in $\Gsmt$ \smtlink{} can
represent the hypothesis \texttt{(booleanp $x_i$)} with the declaration\\
\rule{2em}{0ex}\texttt{x\_i = Bool('x\_i')}\\
without excluding any satisfying assignments.  We assume that the boolean operations
of the SMT solver (e.g. \texttt{And}, \texttt{Or}, \texttt{Not}) correspond
exactly to their ACL2 equivalents when their arguments are boolean.  If a boolean
operator is applied to a non-boolean value, then Z3 throws an exception, and we
regard $G$ as non-translatable.

Similar arguments apply in the case that $x_i$ is an integer, rational, or real
number. We represent ACL2 rational numbers as Z3 real numbers. Because every
rational number is a real number, any satisfying assignment of rational numbers
to rational variables in $\neg \Gtcp$ has a corresponding assignment for $\neg
\Gsmt$. Thus, $\Gsmt$ is a generalization of $\Gtcp$.
We note that for ACL2, formally proving the soundness of this generalization
requires extending our previously discussed $M_2'$ model into a model that
satisfies the theory of real closed
fields~\footnote{http://smtlib.cs.uiowa.edu/theories-Reals.shtml}, because we
are translating rationals in ACL2 to reals in Z3. We haven't wrapped our heads
around how to do that extension in a many-sorted setting, therefore we designate
this to be future work. As with booleans, we assume that arithmetic and
comparison operators have equivalent semantics in ACL2 and the 
SMT solver. In fact, we use the Python interface code to enforce this
assumption. As an example, ACL2 allows the boolean values \texttt{t} and
\texttt{nil} to be used in arithmetic expressions -- both are treated as 0.
Z3 also allows \texttt{True} and \texttt{False} to be used in integer
arithmetic, with \texttt{True} treated as 1 and \texttt{False} treated as 0.
To ensure that $\Gsmt \Rightarrow \Gtcp$, our Python code checks the sorts of the
arguments to arithmetic operators to ensure that they are integers or reals, where
the interpretations are the same for both ACL2 and Z3.

When \smtlink{} is used with ACL2(r), non-classical functions are non-translatable.
We believe that if $\neg \Gtcp$ is classical and satisfiable,
then there exists a satisfying assignment to $\neg \Gtcp$ where all real-valued
variables are bound to standard values. We believe the sketched proof in the
beginning of Section~\ref{sec:theories} works well for ACL2(r). If we are wrong,
we hope that one of the experts in non-standard analysis at the workshop will
correct us.

\subsection{Symbols}
A very important basic type in ACL2 is \texttt{symbolp}. We represent symbols
using an algebraic datatype in Z3. In the z3 interface class, we define a
\texttt{Datatype} called \texttt{z3sym}, with a single field of type
\texttt{IntSort}. Symbol variables are defined using the datatype
\texttt{z3sym}. We then define a class called \texttt{Symbol}. This class
provides a variable \texttt{count} and a variable \texttt{dict}. It also
provides a function called \texttt{intern} for generating a symbol constant.
This class keeps a dictionary mapping from symbol names to the generated
\texttt{z3sym} symbol constants. This creates an injective mapping from symbols
to natural numbers. 
All symbol constants that appeared in the term are mapped onto the first
several, distinct, naturals.

If a satisfying assignment to $\neg \Gtcp$ binds a symbol-valued variable
to a symbol-constant that doesn't appear in $\Gtcp$, then in our soundness
argument, we construct a new symbol value for $\neg \Gsmt$ using an integer
value distinct from the ones used so far -- we won't run out.
Thus, all symbol values in a satisfying assignment to $\neg \Gtcp$
can be translated to corresponding values for $\neg \Gsmt$.
The only operations that we support for symbols are comparisons for equality or
not-equals.  We assume that these operations have corresponding semantics in
ACL2 and the SMT solver.

\subsection{FTY types}
We have added support for common \texttt{fty} types that enable \smtlink{} to
automatically construct bridges from the untyped logic of ACL2 to the typed logic of Z3.
Currently, \smtlink{} infers constructor/destructor relations and other properties
of these types from the \texttt{fty::flextypes-table}.
Thus, the use of \texttt{fty} types extends
the trusted code to include the correctness of these tables.  This trust is mitigated
by two considerations.  First, \smtlink{} only uses \texttt{fty} types that have been
specified by the user in a hint to the \smtlink{} clause processor.  If the user
provides no such hints, no \texttt{fty} types are used by \smtlink{}, and no
soundness concerns arise.

Second, we expect that the information that \smtlink{} obtains from these tables could
be obtained instead from the ACL2 logical world using \texttt{meta-extract} in
\smtlink{}'s verified clause-processor chain.  We see the current implementation
as a useful prototype to explore how to seamlessly infer type information from
code written according to a well-defined type discipline.

\subsubsection{fty::defprod}
The algebraic datatypes of Z3 correspond directly to \texttt{fty::defprod}.
\smtlink{} simply declares a Z3 datatype with a single constructor whose
destructor operators are the field accessors of the product type.
\smtlink{} requires that the arguments to the \texttt{fty} constructors satisfy
the constructors' guards -- otherwise $\Gtcp$ is non-translatable.
The only operations on product types are field accessors, i.e.\ destructors.
For translatable terms, the SMT type has the same construct/destructor theorems
as the FTY type.  Thus, the \smtlink{} translation maintains equivalence
constructors and field accessors of product types.

\subsubsection{fty::deflist}
Lists are essentially a special case of a product type.  For example,\\
\noindent\begin{minipage}[t]{.35\textwidth}\label{prog:deflist}
\begin{lstlisting}[caption=ACL2 deflist,frame=tlrb,style=snippet,language=LISP]{ACL2 deflist}
(deflist integer-list
  :elt-type integerp
  :true-listp t)
\end{lstlisting}
\end{minipage}\hfill
\begin{minipage}[t]{.61\textwidth}
\begin{lstlisting}[caption=Z3 Datatype,frame=tlrb,style=snippet,language=Python]{Z3 Datatype}
integer_list= z3.Datatype('integer_list')
integer_list.declare('cons',
                     ('car', _SMT_.IntSort()),
                     ('cdr', integer_list))
integer_list.declare('nil')
integer_list = integer_list.create()
def integer_list_consp(l):
  return Not(l == integer\_list.nil)
\end{lstlisting}
\end{minipage}
ACL2 overloads \texttt{cons}, \texttt{car}, and \texttt{cdr} to apply to any list.
In contrast, Z3's typed logic has a separate \texttt{cons}, \texttt{car}, \texttt{cdr}
functions for each list type.
This is why our examples from Section~\ref{sec:expl} require fixing functions to convey
the type information to the trusted-clause processor.
We believe that most of the users' burden of typed lists will be removed in a future
release by adding type-inference to \smtlink{}. There are soundness issues that
must be addressed. In ACL2, \texttt{(equal (car nil) nil)}.
In Z3,\\
\rule{2em}{0ex}\texttt{integer\_list.car(integer\_list.nil)}\\
is an arbitrary integer.
To ensure soundness, the trusted-clause processor produces
the proof obligation \texttt{(consp x)} for every occurrence of \texttt{(car x)}
that it encounters.
Under this precondition, the Z3 translation preserves the constructor/destructor
relationship for lists.
Likewise\\
\rule{2em}{0ex}\texttt{integer\_list.cdr(integer\_list.nil)}\\
is an arbitrary \texttt{integer\_list}.
Thus, \smtlink{} enforces the \texttt{:true-listp t} declaration for list types.
Because ``arbitrary'' includes \texttt{integer\_list.nil} in addition to all other \texttt{integer\_list}s,
these construction
ensures that the SMT solver can choose the value for \texttt{integer\_list.cdr($\tilde{\texttt{x}}$)} for $\Gsmt$
that corresponds to the value of \texttt{(cdr x)} for any assignments for $\Gtcp$.
The $\Gsmt$ is a generalization of $\Gtcp$.

\subsubsection{fty::defalist}
\smtlink{} represents alists with SMT arrays.  We only support the operations
\texttt{acons} and \texttt{assoc-equal} for alists.  Then we have:
\begin{lstlisting}[style=snippet]
(defthm alist-axioms
  (implies (not (equal key1 key2))
           (and (equal (assoc key1 (acons key1 value alist)) value))
                (equal (assoc key1 (acons key2 value alist)) (assoc key1 alist)))
                (equal (assoc key1 nil) nil))
\end{lstlisting}
The corresponding theorem in the theory of arrays is
\begin{lstlisting}[style=snippet]
(defthm array-axioms
  (implies (not (equal addr1 addr2))
           (and (equal (load addr1 (store addr1 value array)) value))
                (equal (load addr1 (store addr2 value array)) (load addr1 array))))
\end{lstlisting}
Note that \smtlink{} does not support operations such as \texttt{cdr},
\texttt{nth}, \texttt{member}, or \texttt{delete-assoc} that would
``remove'' elements from an alist. Also, Z3 arrays are typed.

The key issue in the translation is how to handle the case when a key is not found in the
alist (ACL2) or array (SMT).
Our solution is to make the element type of the SMT array be an option type called
\texttt{keyType\_elementType}.  This type can either be a \texttt{(key, value)} tuple
or \texttt{keyType\_elementType.nil}.
Thus, any value returned by \texttt{assoc-equal} with proper alist and key types
has a corresponding \texttt{keyType\_elementType} value.  Thus, any value for an
\texttt{assoc-equal} terms in $\Gtcp$ can be represented in $\Gsmt$.

When applying \texttt{cdr} to a \texttt{keyType\_elementType}, we must ensure that the
\texttt{keyType\_elementType} value is not nil.  This is analogous to the issue with lists:
in ACL2, \texttt{(equal (cdr nil) nil)} but in Z3,\\
\rule{2em}{0ex}\texttt{keyType\_elementType.cdr(keyType\_elementType.nil)}\\
is an arbitrary value of \texttt{elementType}.
Thus, the trusted-clause processor produces the proof obligation for ACL2
\texttt{(not (null x))} for every occurrence of \texttt{(cdr x)} when \texttt{x} is the
return value from \texttt{assoc-equal}.
Under this precondition, \texttt{cdr} is only applied to non-nil values from
\texttt{assoc-equal} and we maintain correspondence of values for terms in
$\Gtcp$ and $\Gsmt$. 

By only providing \texttt{acons} and \texttt{assoc-equal}, the \smtlink{}
support for alists is rather limited.  Nevertheless, we have found it to be very
useful when reasoning about problems where alists are used as simple
dictionaries. 

\subsubsection{fty::defoption}
As is shown in Program~\ref{prog:defoption}, the translation of
\texttt{defoption} is straightforward.

\noindent\begin{minipage}{.35\textwidth}\label{prog:defoption}
\begin{lstlisting}[caption=ACL2 deflist,frame=tlrb,style=snippet,language=LISP]{ACL2 defoption}
(defoption maybe-integer
           integerp)


\end{lstlisting}
\end{minipage}\hfill
\begin{minipage}{.61\textwidth}
\begin{lstlisting}[caption=Z3 Datatype,frame=tlrb,style=snippet,language=Python]{Z3 Datatype}
maybe_integer= z3.Datatype('maybe_integer')
maybe_integer.declare('some', ('val', IntSort()))
maybe_integer.declare('nil')
maybe_integer = maybe_integer.create()
\end{lstlisting}
\end{minipage}

In this example, the \texttt{maybe-integer-p} recognizer maps to
\texttt{maybe_integer} type. The constructor \texttt{maybe-integer-some} maps to
\texttt{maybe_integer.some}. The destructor \texttt{maybe-integer-some->val} 
maps to \texttt{maybe_integer.val}. The none type \texttt{nil} maps to
\texttt{maybe_integer.nil}. Typical users of FTY types won't write
\texttt{maybe-integer-some} constructor and \texttt{maybe-integer-some->val}
destructors. They will first check if a term is nil, and then assume the term is
an \texttt{integerp}. When a program returns an \texttt{integerp}, ACL2 knows it
is also a \texttt{maybe-integerp}. Due to lack of type inference capabilities,
\smtlink{} currently requires the user to use those constructors, therefore
maintaining the option type through function calls where a
\texttt{maybe-integerp} is needed and use the destructors where an
\texttt{integerp} is needed.  The constructor function satisfies the same theorems
in ACL2 and Z3.  Therefore, it's sound.  For the field-accessor, when trying to
access field of a \texttt{nil}, ACL2 returns the fixed default value, while Z3
will return arbitrary value of that \texttt{some} type. The Z3 values include
the ACL2 value.  \texttt{nil} is trivially the same. Thus, the \smtlink{}
translation maintains equivalence of values of terms for constructors and field
accessors of option types.

\subsection{Uninterpreted Functions}
The user can direct \smtlink{} to represent some functions as uninterpreted functions
in the SMT theories.  Let \texttt{(f arg1 arg2 \ldots argk)} be a function instance
in $\Gtcp$.  \smtlink translates this to\\
\rule{2em}{0ex}\texttt{f\_smt(arg\_smt1, arg\_smt2, \ldots, arg\_smtk)}\\
The constraints for \texttt{f\_smt} are the types of the argument, the type of the
result, and any user-specified constraints.
If the function instance in $\Gtcp$ violates the argument type constraints, then
the term is untranslatable -- currently, \smtlink{} produces an SMT term
that provokes a \texttt{z3types.Z3exception}.
For each function instance in $\Gtcp$ that is translated to an uninterpreted function,
\smtlink{} produces a proof obligation for ACL2 that the function instances in
$\Gtcp$ satisfy the given type-recognizers.  Likewise, if the user specified
any other constraints for this function, they are returned as further ACL2 proof
obligations.
Under these preconditions, any value that can be produced by \texttt{f} satisfies the
constraints for \texttt{f\_smt}.
Thus, we maintain correspondence of values for terms in $\Gtcp$ and $\Gsmt$.


\section{A Ring Oscillator Example}\label{sec:expl}
\begin{figure}[h]
  \begin{center}
   	\includegraphics[width=0.5\textwidth]{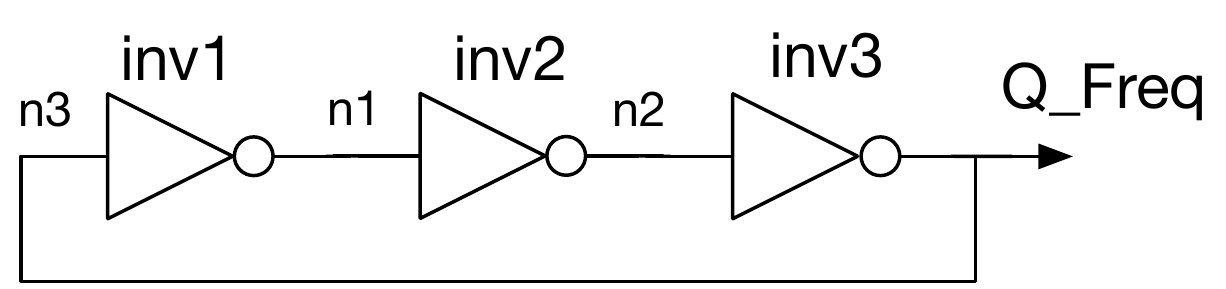}
    \caption[3-Stage Ring Oscillator]{\label{fig:ring} 3-Stage Ring Oscillator}
  \end{center}
\end{figure}

In order to show the power of \smtlink{}, especially what benefit FTY types
bring us, in this section, we take a look at how \smtlink{} can be used in
proving an invariant of a small circuit. This example uses extensively the FTY
types, including product types, list types, alist types, and option types. The
simple circuit we want to model is a 3-stage ring oscillator. A ring oscillator
is an oscillator circuit consisting of an odd number of inverters in a ring as
is shown in Figure~\ref{fig:ring}. A 3-stage ring oscillator consists of three
inverters. It oscillates to provide a signal of a certain frequency.

For each inverter in this circuit, we say it is stable if its input is not equal
to its output, otherwise, we say the inverter is ready-to-fire. One interesting
invariant of this circuit is defined in Theorem~\ref{thm:ring}:

\begin{theorem}\label{thm:ring}
  Starting from a state where there is one and only one inverter ready-to-fire,
  for all future states, the ring oscillator will stay in a state where there is
  only one inverter ready-to-fire.
\end{theorem}

In order to prove this property of this ring oscillator, we will discuss how we
used FTY types extensively for circuit modeling and how \smtlink{} helped
greatly at proving the theorem. To check the details of the proof, see
\href{https://github.com/acl2/acl2/tree/master/books/projects/smtlink/examples}{projects/smtlink/examples/ringosc.lisp}

\subsection{Circuit and Trace Modeling using FTY Types}
We model an \textbf{inverter} gate by defining a product type with two fields --
\texttt{input} and \texttt{output}. We then model the 3-stage \textbf{ring
  oscillator} using a product type with six fields -- three internal nodes
\texttt{n1} through \texttt{n3} and three submodule inverters \texttt{inv1}
through \texttt{inv3}. We call it \texttt{ringosc3-p}. We then define a
\textbf{connection function} specifying the connections between the upper-level
ring oscillator nodes and the lower-level inverter nodes. This describes the
shape of the circuit.

As for modeling the behavior of the circuit, we use traces~\cite{Dill87}. We
define a circuit \textbf{state} as an alist mapping from signal names to its
values. A \textbf{trace} is a list of circuits states, called
\texttt{any-trace-p}. We define a \textbf{step recognizer} for an inverter. This
recognizer function takes two consecutive steps from a trace as inputs and
serves as a constraint function of what are the allowed behaviors in a step for
an inverter. A valid trace for an inverter is defined recursively using the step
function. A valid trace for the 3-stage ring oscillator is then defined
requiring the trace to be valid for all three inverters. We call the recognizer
function for a valid trace of a ring oscillator, \texttt{ringosc3-valid}.

We define a \textbf{counting} function for an inverter. The counting function
returns $1$ when an inverter is ready-to-fire and $0$ when it's stable. Then we
define a counting function for the ring oscillator based on the counting
function for an inverter. We say a state of the ring oscillator is
\textit{one-safe} if only one inverter is ready-to-fire. We call the function
\texttt{ringosc3-one-safe-state}. Using this function, we can define
\texttt{ringosc3-one-safe-trace}, which means all states in a trace are
\textit{one-safe}. Theorem~\ref{thm:ring} is defined as Program~\ref{prog:ring}.

\begin{Program}[h]
  \caption{The ringosc3-one-safe theorem}
  \label{prog:ring}
  \scalebox{1.0}{\begin{minipage}{1.00\textwidth}
\begin{lstlisting}[language=Lisp]
(defthm ringosc3-one-safe
  (implies (and (ringosc3-p r)
                (any-trace-p tr)
                (consp tr)
                (ringosc3-valid r tr)
                (ringosc3-one-safe-state r (car tr)))
           (ringosc3-one-safe-trace r tr))
  :hints (("Goal"
           :induct (ringosc3-one-safe-trace r tr)
           :in-theory (e/d (ringosc3-one-safe-trace
                            ringosc3-valid
                            inverter-valid)
                           (ringosc3-one-safe-lemma)))
          ("Subgoal *1/1.1"
           :use ((:instance ringosc3-one-safe-lemma
                            (r r)
                            (tr tr)))
           )))
\end{lstlisting}\end{minipage}}
\end{Program}

In this theorem, \smtlink{} helped to prove the inductive step. Due to space constraints,
the details of \texttt{ringosc3-one-safe-lemma} are elided in this paper.. We note
that proving this theorem using just ACL2 requires proving detailed lemmas about
possible transitions of the ring oscillator. More specifically, our trace
model requires asking if a signal exists in the state table, which causes a huge
amount of case splits. However, this has not been a problem for \smtlink{} and
the large amount of cases is handled efficiently.
Using \smtlink{}, we are able to expand the functions out to proper steps and
then prove the theorem without much human intervention. This demonstrates the
potential of applying \smtlink{} to systems and proofs about systems. All this
is made convenient because of the new theories supported and the useful
user-interface.

\section{Related Work}\label{sec:rlwk}

Integrating external procedures like SAT and SMT solvers into ACL2 has been done
in several works in the past. Srinivasan \cite{srinivasan2007} integrated
the Yices~\cite{Dutertre2014} SMT solver into ACL2 for verifying bit-level
pipelined machines.  They use a trusted clause processor with a translation
process.  They appear to have mostly used the bit-vector arithmetic and SAT
solving capabilities of Yices. Prior to that, in~\cite{Manolios2006}, they
integrated a decision procedure called UCLID~\cite{Lahiri2004} into ACL2 to
solve a similar problem. These are works that require fully trusting the
integration.

A typical way of ensuring soundness and avoiding trusting external procedures is
to followed Harrison and Th\'{e}ry's ``skeptical'' approach~\cite{Harrison98}
and reconstruct proofs in the theorem prover. Recent work by
Lu{\'i}s~\cite{Cruz2017} allows refutation proofs of SAT problems to be
reconstructed and replayed inside of ACL2. Their work focused on generating
efficient refutation proofs that can be checked by a theorem prover in a short
amount of time. Integrating SMT solvers into theorem provers has been a
consistently developing area in the past
decade~\cite{Mclaughlin06,Fontaine06,Besson06,Merz12,Deharbe14,Blanchette13,Armand11,Erkok08}.
Erk{\"{o}}k~\cite{Erkok08} integrated the SMT solver Yices into Isabelle/HOL.
Similar to \smtlink{}, they not only have basic theories but also support
algebraic datatypes. They trust Yices as an oracle.
Works like ~\cite{Blanchette13,Burak17} do proof reconstruction.
Sledgehammer~\cite{Blanchette13, Blanchette10} is a proof assistant that
integrates a bunch of SMT solvers into the theorem prover Isabelle/HOL. Proof
reconstruction task is hard, and as pointed out in~\cite{Blanchette10} the
reconstruction can fail, and sometimes take a tremendous amount of time.
Armand~\cite{Armand11,Burak17} developed a framework in the theorem prover Coq
for integrating external SMT solver results. They developed a set of ``small
checkers'' that are able to take a certificate and call corresponding small
checker for proof checks and used Coq tactics for automation. They report having
achieved better performance than~\cite{Blanchette13}. Also working on bridging
the gap between interactive theorem proving and automated theorem proving (aka
solvers), Moura~\cite{Moura15} chooses a different path to build a theorem
prover called Lean which also uses Z3, the SMT solver, as a back-end, but also
provides benefits of an interactive theorem prover.

Several papers showed how their methods could be used for the verification of
concurrent algorithms such as clock synchronization \cite{Fontaine06}, and the
Bakery and Memoir algorithms \cite{Merz12}. Erk{\"o}k~\cite{Erkok08} uses the
integration to prove memory safety of small C programs. \cite{Fontaine06} used
the CVC-Lite \cite{Barrett04} SMT solver to verify properties of simple
quadratic inequalities. SMT solvers have drawn interests in the programming
language research where ~\cite{Vazou:2017} integrates SMT solvers into Haskell
for proving refinement types.

Our work is based on our previous work~\cite{Peng2015-acl2}. Previously we showed how
the integration of SMT solvers with theorem provers can help to prove properties
of Analog/Mixed-Signal Designs. We used a single trusted clause-processor like
is done in~\cite{srinivasan2007}. This paper describes our recent work of
re-architecting \smtlink{} to verify the majority of it using verified
clause-processors, therefore greatly improved soundness. We now depend on a very
minimal core in a trusted clause-processor. In addition to that, our added
support for algebraic datatypes largely broadened the horizon of problems
\smtlink{} is able to handle.

\section{Conclusion and Future Work}\label{sec:concl}

In this paper, we discussed an updated \smtlink{}. Comparing to the previous
version, the current \smtlink{} has a more compelling argument of soundness,
is extensible, and supports more theories. The architecture of \smtlink{} now
clearly separates into verified and trusted parts. We make the trusted core as
small as possible. We outlined an approach for verifying soundness, explaining
how the gap is bridged between logic of ACL2 and the logic of an SMT solver.
The new architecture through a sequence of verified clause-processors makes it
extensible and easy to maintain. There are still many aspects we want to keep
working on.

First, we want to use the meta-extract~\cite{KaufmannS17} capability introduced
in last year's workshop. We believe if we can use the meta-extract to fully
verify several of our clause-processors, for example, the function expansion
clause processor, then the clause-processor won't need to return the clause back
to ACL2 for proof. We observe that when projects get larger, the auxiliary
clauses can become hard to prove, and making the clause-processor fully verified
will reduce time spent on proving the auxiliary clauses to 0.

Second, given that we have the meta-extract capability, we are wondering if we
can make a verified type inference engine. Currently, \smtlink{} knows nothing
about types of terms and replies on the user for type instrumentation. Fixing
functions have to be added to places where such a type information is required.
For example, when dealing with a \texttt{nil}, which type of \texttt{nil} is it?
We would love to deduce the types of terms and relieve the burden on the users
for specifying types.

Third, we sketched a soundness proof in this paper, but this proof is not
complete. Specifically we need to extend the model $M_2'$ as described in
Section~\ref{sec:theories} to a model that satisfies the real-closure
axioms in a many-sorted setting. The current soundness proof sketch works well
with ACL2(r) but not ACL2. We'd like to complete this proof.

Fourth, we want to make all counter-examples evaluable. This will require knowing
FTY types and how to translate their constructors back. We also have to figure
out how to translate algebraic numbers.

Fifth, we note \texttt{Satlink} has implemented a proof reconstruction
interface that allows proofs to be returned from an SAT solver and replayed
in ACL2. This can make the single trusted clause-processor goes away and remove
all trusts we give to external SMT solvers. Proof reconstruction is an
interesting direction that we might want to research more.

For applications, we believe the current \smtlink{} have enough capability that
it can be applied to a lot of problems. We have been working on using it to
verify properties of an asynchronous FIFO. The results are promising. In the
future, we want to use it to prove timing properties making use of its
arithmetic reasoning ability.

\section*{Acknowledgments}
We would like to thank the ACL2 community for all the help in answering our
questions while using ACL2 and developing \smtlink{}. We especially want to
thank Matt Kaufmann for teaching us model theory and the many discussions
to form a sketch of the soundness proof for \smtlink{}. We are also thankful to
the anonymous reviewers for the insightful and constructive feedback.

\nocite{*}
\bibliographystyle{eptcs}
\bibliography{ref}

\begin{thebibliography}{10}
\providecommand{\bibitemdeclare}[2]{}
\providecommand{\surnamestart}{}
\providecommand{\surnameend}{}
\providecommand{\urlprefix}{Available at }
\providecommand{\url}[1]{\texttt{#1}}
\providecommand{\href}[2]{\texttt{#2}}
\providecommand{\urlalt}[2]{\href{#1}{#2}}
\providecommand{\doi}[1]{doi:\urlalt{http://dx.doi.org/#1}{#1}}
\providecommand{\bibinfo}[2]{#2}

\bibitemdeclare{inproceedings}{Armand11}
\bibitem{Armand11}
\bibinfo{author}{M.~\surnamestart Armand\surnameend},
  \bibinfo{author}{G.~\surnamestart Faure\surnameend},
  \bibinfo{author}{B.~\surnamestart Gr{\'e}goire\surnameend},
  \bibinfo{author}{C.~\surnamestart Keller\surnameend},
  \bibinfo{author}{L.~\surnamestart Th{\'e}ry\surnameend} \&
  \bibinfo{author}{B.~\surnamestart Werner\surnameend} (\bibinfo{year}{2011}):
  \emph{\bibinfo{title}{A Modular Integration of SAT/SMT Solvers to Coq Through
  Proof Witnesses}}.
\newblock In: {\sl \bibinfo{booktitle}{1st Int'l. Conf. Certified Programs and
  Proofs}}, \bibinfo{publisher}{Springer}, pp. \bibinfo{pages}{135--150},
  \doi{10.1007/978-3-642-25379-9\_12}.

\bibitemdeclare{incollection}{Barrett04}
\bibitem{Barrett04}
\bibinfo{author}{C.~\surnamestart Barrett\surnameend} \&
  \bibinfo{author}{S.~\surnamestart Berezin\surnameend} (\bibinfo{year}{2004}):
  \emph{\bibinfo{title}{CVC Lite: A New Implementation of the Cooperating
  Validity Checker}}.
\newblock In: {\sl \bibinfo{booktitle}{Computer Aided Verification}}, {\sl
  \bibinfo{series}{LNCS}} \bibinfo{volume}{3114},
  \bibinfo{publisher}{Springer}, pp. \bibinfo{pages}{515--518},
  \doi{10.1007/978-3-540-27813-9\_49}.

\bibitemdeclare{misc}{Barrett10}
\bibitem{Barrett10}
\bibinfo{author}{C.~\surnamestart Barrett\surnameend},
  \bibinfo{author}{A.~\surnamestart Stump\surnameend} \&
  \bibinfo{author}{C.~\surnamestart Tinelli\surnameend} (\bibinfo{year}{2010}):
  \emph{\bibinfo{title}{{The SMT-LIB Standard: Version 2.0}}}.
\newblock
  \bibinfo{howpublished}{\url{http://www.cs.nyu.edu/~barrett/pubs/BST10.pdf}}.
\newblock \bibinfo{note}{[Online; accessed 17-August-2015]}.

\bibitemdeclare{inproceedings}{Besson06}
\bibitem{Besson06}
\bibinfo{author}{F.~\surnamestart Besson\surnameend} (\bibinfo{year}{2007}):
  \emph{\bibinfo{title}{Fast Reflexive Arithmetic Tactics the Linear Case and
  Beyond}}.
\newblock In: {\sl \bibinfo{booktitle}{2006 Int'l. Conf. Types for Proofs and
  Programs}}, \bibinfo{publisher}{Springer}, pp. \bibinfo{pages}{48--62},
  \doi{10.1007/978-3-540-74464-1\_4}.

\bibitemdeclare{article}{Blanchette13}
\bibitem{Blanchette13}
\bibinfo{author}{J.C. \surnamestart Blanchette\surnameend},
  \bibinfo{author}{S.~\surnamestart B\"{o}hme\surnameend} \&
  \bibinfo{author}{L.C. \surnamestart Paulson\surnameend}
  (\bibinfo{year}{2013}): \emph{\bibinfo{title}{Extending Sledgehammer with
  {SMT} Solvers}}.
\newblock {\sl \bibinfo{journal}{J. Automated Reasoning}}
  \bibinfo{volume}{51}(\bibinfo{number}{1}), pp. \bibinfo{pages}{109--128},
  \doi{10.1007/s10817-013-9278-5}.

\bibitemdeclare{article}{Deharbe14}
\bibitem{Deharbe14}
\bibinfo{author}{D.~\surnamestart D{\'e}harbe\surnameend},
  \bibinfo{author}{P.~\surnamestart Fontaine\surnameend},
  \bibinfo{author}{Y.~\surnamestart Guyot\surnameend} \&
  \bibinfo{author}{L.~\surnamestart Voisin\surnameend} (\bibinfo{year}{2014}):
  \emph{\bibinfo{title}{Integrating {SMT} Solvers in Rodin}}.
\newblock {\sl \bibinfo{journal}{Sci. Comput. Program.}}
  \bibinfo{volume}{94}(\bibinfo{number}{P2}), pp. \bibinfo{pages}{130--143},
  \doi{10.1016/j.scico.2014.04.012}.

\bibitemdeclare{phdthesis}{Dill87}
\bibitem{Dill87}
\bibinfo{author}{David~L. \surnamestart Dill\surnameend}
  (\bibinfo{year}{1987}): \emph{\bibinfo{title}{Trace Theory for Automatic
  Hierarchical Verification of Speed-independent Circuits}}.
\newblock Ph.D. thesis, \bibinfo{school}{Carnegie Mellon University},
  \bibinfo{address}{Pittsburgh, PA, USA}.
\newblock
  \urlprefix\url{http://reports-archive.adm.cs.cmu.edu/anon/scan/CMU-CS-88-119.pdf}.
\newblock \bibinfo{note}{AAI8814716}.

\bibitemdeclare{incollection}{Dutertre2014}
\bibitem{Dutertre2014}
\bibinfo{author}{B.~\surnamestart Dutertre\surnameend} (\bibinfo{year}{2014}):
  \emph{\bibinfo{title}{Yices 2.2}}.
\newblock In: {\sl \bibinfo{booktitle}{Computer Aided Verification}}, {\sl
  \bibinfo{series}{LNCS}} \bibinfo{volume}{8559},
  \bibinfo{publisher}{Springer}, pp. \bibinfo{pages}{737--744},
  \doi{10.1007/978-3-319-08867-9\_49}.

\bibitemdeclare{inproceedings}{Burak17}
\bibitem{Burak17}
\bibinfo{author}{Burak \surnamestart Ekici\surnameend}, \bibinfo{author}{Alain
  \surnamestart Mebsout\surnameend}, \bibinfo{author}{Cesare \surnamestart
  Tinelli\surnameend}, \bibinfo{author}{Chantal \surnamestart
  Keller\surnameend}, \bibinfo{author}{Guy \surnamestart Katz\surnameend},
  \bibinfo{author}{Andrew \surnamestart Reynolds\surnameend} \&
  \bibinfo{author}{Clark \surnamestart Barrett\surnameend}
  (\bibinfo{year}{2017}): \emph{\bibinfo{title}{SMTCoq: A Plug-In for
  Integrating SMT Solvers into Coq}}.
\newblock In \bibinfo{editor}{Rupak \surnamestart Majumdar\surnameend} \&
  \bibinfo{editor}{Viktor \surnamestart Kun{\v{c}}ak\surnameend}, editors: {\sl
  \bibinfo{booktitle}{Computer Aided Verification}},
  \bibinfo{publisher}{Springer International Publishing},
  \bibinfo{address}{Cham}, pp. \bibinfo{pages}{126--133},
  \doi{10.1007/978-3-319-63390-9_7}.

\bibitemdeclare{inproceedings}{Erkok08}
\bibitem{Erkok08}
\bibinfo{author}{Levent \surnamestart Erk{\"{o}}k\surnameend} \&
  \bibinfo{author}{John \surnamestart Matthews\surnameend}
  (\bibinfo{year}{2008}): \emph{\bibinfo{title}{Using Yices as an Automated
  Solver in Isabelle/HOL}}.
\newblock In: {\sl \bibinfo{booktitle}{In Automated Formal Methods’08}},
  \bibinfo{publisher}{ACM Press}, pp. \bibinfo{pages}{3--13}.
\newblock
  \urlprefix\url{http://citeseerx.ist.psu.edu/viewdoc/summary?doi=10.1.1.156.8123}.

\bibitemdeclare{inproceedings}{Fontaine06}
\bibitem{Fontaine06}
\bibinfo{author}{P.~\surnamestart Fontaine\surnameend}, \bibinfo{author}{J.-Y.
  \surnamestart Marion\surnameend}, \bibinfo{author}{S.~\surnamestart
  Merz\surnameend}, \bibinfo{author}{L.P. \surnamestart Nieto\surnameend} \&
  \bibinfo{author}{A.~\surnamestart Tiu\surnameend} (\bibinfo{year}{2006}):
  \emph{\bibinfo{title}{Expressiveness + Automation + Soundness: Towards
  Combining SMT Solvers and Interactive Proof Assistants}}.
\newblock In: {\sl \bibinfo{booktitle}{12th Int'l. Conf. Tools and Algorithms
  for the Construction and Analysis of Systems}},
  \bibinfo{publisher}{Springer}, pp. \bibinfo{pages}{167--181},
  \doi{10.1007/11691372\_11}.

\bibitemdeclare{article}{Harrison98}
\bibitem{Harrison98}
\bibinfo{author}{J.~\surnamestart Harrison\surnameend} \&
  \bibinfo{author}{L.~\surnamestart Th{\'e}ry\surnameend}
  (\bibinfo{year}{1998}): \emph{\bibinfo{title}{A Skeptic's Approach to
  Combining HOL and Maple}}.
\newblock {\sl \bibinfo{journal}{J. Automated Reasoning}}
  \bibinfo{volume}{21}(\bibinfo{number}{3}), pp. \bibinfo{pages}{279--294},
  \doi{10.1023/A:1006023127567}.

\bibitemdeclare{inproceedings}{Cruz2017}
\bibitem{Cruz2017}
\bibinfo{author}{Marijn \surnamestart Heule\surnameend},
  \bibinfo{author}{Warren \surnamestart Hunt\surnameend}, \bibinfo{author}{Matt
  \surnamestart Kaufmann\surnameend} \& \bibinfo{author}{Nathan \surnamestart
  Wetzler\surnameend} (\bibinfo{year}{2017}): \emph{\bibinfo{title}{Efficient,
  Verified Checking of Propositional Proofs}}.
\newblock In \bibinfo{editor}{Mauricio \surnamestart
  Ayala-Rinc{\'o}n\surnameend} \& \bibinfo{editor}{C{\'e}sar~A. \surnamestart
  Mu{\~{n}}oz\surnameend}, editors: {\sl \bibinfo{booktitle}{Interactive
  Theorem Proving}}, \bibinfo{publisher}{Springer International Publishing},
  \bibinfo{address}{Cham}, pp. \bibinfo{pages}{269--284},
  \doi{10.1007/978-3-319-66107-0_18}.

\bibitemdeclare{misc}{Blanchette10}
\bibitem{Blanchette10}
\bibinfo{author}{L.~Paulson \surnamestart J.~Blanchette\surnameend}
  (\bibinfo{year}{2017}): \emph{\bibinfo{title}{{Sledgehammer}}}.
\newblock
  \bibinfo{howpublished}{\url{https://isabelle.in.tum.de/dist/doc/sledgehammer.pdf}}.
\newblock \bibinfo{note}{[Online; accessed 14-July-2018]}.

\bibitemdeclare{inproceedings}{KaufmannS17}
\bibitem{KaufmannS17}
\bibinfo{author}{Matt \surnamestart Kaufmann\surnameend} \&
  \bibinfo{author}{Sol \surnamestart Swords\surnameend} (\bibinfo{year}{2017}):
  \emph{\bibinfo{title}{Meta-extract: Using Existing Facts in Meta-reasoning}}.
\newblock In \bibinfo{editor}{Slobodov{\'{a}}} \& \bibinfo{editor}{Jr.}
  \cite{DBLP:journals/corr/SlobodovaJ17}, pp. \bibinfo{pages}{47--60},
  \doi{10.4204/EPTCS.249.4}.
\newblock \urlprefix\url{http://arxiv.org/abs/1705.00766}.

\bibitemdeclare{incollection}{Lahiri2004}
\bibitem{Lahiri2004}
\bibinfo{author}{S.K. \surnamestart Lahiri\surnameend} \& \bibinfo{author}{S.A.
  \surnamestart Seshia\surnameend} (\bibinfo{year}{2004}):
  \emph{\bibinfo{title}{The UCLID Decision Procedure}}.
\newblock In: {\sl \bibinfo{booktitle}{Computer Aided Verification}}, {\sl
  \bibinfo{series}{LNCS}} \bibinfo{volume}{3114},
  \bibinfo{publisher}{Springer}, pp. \bibinfo{pages}{475--478},
  \doi{10.1007/978-3-540-27813-9\_40}.

\bibitemdeclare{article}{Manolios2006}
\bibitem{Manolios2006}
\bibinfo{author}{P.~\surnamestart Manolios\surnameend} \& \bibinfo{author}{S.K.
  \surnamestart Srinivasan\surnameend} (\bibinfo{year}{2006}):
  \emph{\bibinfo{title}{A Framework for Verifying Bit-Level Pipelined Machines
  Based on Automated Deduction and Decision Procedures}}.
\newblock {\sl \bibinfo{journal}{J. of Automated Reasoning}}
  \bibinfo{volume}{37}(\bibinfo{number}{1-2}), pp. \bibinfo{pages}{93--116},
  \doi{10.1007/s10817-006-9035-0}.

\bibitemdeclare{inproceedings}{Mclaughlin06}
\bibitem{Mclaughlin06}
\bibinfo{author}{S.~\surnamestart Mclaughlin\surnameend}, \bibinfo{author}{Cl.
  \surnamestart Barrett\surnameend} \& \bibinfo{author}{Y.~\surnamestart
  Ge\surnameend} (\bibinfo{year}{2006}): \emph{\bibinfo{title}{Cooperating
  Theorem Provers: A Case Study Combining HOL-Light and CVC Lite}}.
\newblock In: {\sl \bibinfo{booktitle}{In Proc. 3rd Workshop on Pragmatics of
  Decision Procedures in Automated Reasoning}}, {\sl \bibinfo{series}{ENTCS}}
  \bibinfo{volume}{144(2)}, \bibinfo{publisher}{Elsevier}, pp.
  \bibinfo{pages}{43--51}, \doi{10.1016/j.entcs.2005.12.005}.

\bibitemdeclare{inproceedings}{Merz12}
\bibitem{Merz12}
\bibinfo{author}{S.~\surnamestart Merz\surnameend} \&
  \bibinfo{author}{H.~\surnamestart Vanzetto\surnameend}
  (\bibinfo{year}{2012}): \emph{\bibinfo{title}{Automatic Verification of
  TLA$^{+}$; Proof Obligations with SMT Solvers}}.
\newblock In: {\sl \bibinfo{booktitle}{18th Int'l. Conf. Logic for Programming,
  Artificial Intelligence, and Reasoning}}, \bibinfo{publisher}{Springer}, pp.
  \bibinfo{pages}{289--303}, \doi{10.1007/978-3-642-28717-6\_23}.

\bibitemdeclare{inproceedings}{Moura08}
\bibitem{Moura08}
\bibinfo{author}{Leonardo \surnamestart Moura\surnameend} \&
  \bibinfo{author}{Nikolaj \surnamestart Bj{\o}rner\surnameend}
  (\bibinfo{year}{2008}): \emph{\bibinfo{title}{Z3: {An} Efficient {SMT}
  Solver}}.
\newblock In \bibinfo{editor}{C.R. \surnamestart Ramakrishnan\surnameend} \&
  \bibinfo{editor}{Jakob \surnamestart Rehof\surnameend}, editors: {\sl
  \bibinfo{booktitle}{Tools and Algorithms for the Construction and Analysis of
  Systems}}, {\sl \bibinfo{series}{Lecture Notes in Computer Science}}
  \bibinfo{volume}{4963}, \bibinfo{publisher}{Springer Berlin Heidelberg}, pp.
  \bibinfo{pages}{337--340}, \doi{10.1007/978-3-540-78800-3_24}.

\bibitemdeclare{inproceedings}{Moura15}
\bibitem{Moura15}
\bibinfo{author}{Leonardo \surnamestart de~Moura\surnameend},
  \bibinfo{author}{Soonho \surnamestart Kong\surnameend},
  \bibinfo{author}{Jeremy \surnamestart Avigad\surnameend},
  \bibinfo{author}{Floris \surnamestart van Doorn\surnameend} \&
  \bibinfo{author}{Jakob \surnamestart von Raumer\surnameend}
  (\bibinfo{year}{2015}): \emph{\bibinfo{title}{The Lean Theorem Prover (System
  Description)}}.
\newblock In \bibinfo{editor}{Amy~P. \surnamestart Felty\surnameend} \&
  \bibinfo{editor}{Aart \surnamestart Middeldorp\surnameend}, editors: {\sl
  \bibinfo{booktitle}{Automated Deduction - CADE-25}},
  \bibinfo{publisher}{Springer International Publishing}, pp.
  \bibinfo{pages}{378--388}, \doi{10.1007/978-3-319-21401-6_26}.

\bibitemdeclare{inproceedings}{Peng2015-acl2}
\bibitem{Peng2015-acl2}
\bibinfo{author}{Yan \surnamestart Peng\surnameend} \& \bibinfo{author}{Mark
  \surnamestart Greenstreet\surnameend} (\bibinfo{year}{2015}):
  \emph{\bibinfo{title}{Extending {ACL2} with {SMT} Solvers}}.
\newblock In \bibinfo{editor}{Matt \surnamestart Kaufmann\surnameend} \&
  \bibinfo{editor}{David~L. \surnamestart Rager\surnameend}, editors: {\sl
  \bibinfo{booktitle}{Proceedings Thirteenth International Workshop on the
  {ACL2} Theorem Prover and Its Applications}}, {\sl
  \bibinfo{series}{Electronic Proceedings in Theoretical Computer Science}}
  \bibinfo{volume}{192}, \bibinfo{publisher}{Open Publishing Association},
  \bibinfo{address}{Austin, Texas, USA, 1--2 October 2015}, pp.
  \bibinfo{pages}{61--77}, \doi{10.4204/EPTCS.192.6}.

\bibitemdeclare{inproceedings}{Peng2015-nfs}
\bibitem{Peng2015-nfs}
\bibinfo{author}{Yan \surnamestart Peng\surnameend} \& \bibinfo{author}{Mark
  \surnamestart Greenstreet\surnameend} (\bibinfo{year}{2015}):
  \emph{\bibinfo{title}{Integrating {SMT} with Theorem Proving for
  Analog/Mixed-Signal Circuit Verification}}.
\newblock In \bibinfo{editor}{Klaus \surnamestart Havelund\surnameend},
  \bibinfo{editor}{Gerard \surnamestart Holzmann\surnameend} \&
  \bibinfo{editor}{Rajeev \surnamestart Joshi\surnameend}, editors: {\sl
  \bibinfo{booktitle}{NASA Formal Methods}}, \bibinfo{publisher}{Springer
  International Publishing}, pp. \bibinfo{pages}{310--326},
  \doi{10.1007/978-3-319-17524-9_22}.

\bibitemdeclare{proceedings}{DBLP:journals/corr/SlobodovaJ17}
\bibitem{DBLP:journals/corr/SlobodovaJ17}
\bibinfo{editor}{Anna \surnamestart Slobodov{\'{a}}\surnameend} \&
  \bibinfo{editor}{Warren A.~Hunt \surnamestart Jr.\surnameend}, editors
  (\bibinfo{year}{2017}): \emph{\bibinfo{title}{Proceedings 14th International
  Workshop on the {ACL2} Theorem Prover and its Applications, Austin, Texas,
  USA, May 22-23, 2017}}. {\sl \bibinfo{series}{{EPTCS}}}
  \bibinfo{volume}{249}.
\newblock \urlprefix\url{http://arxiv.org/abs/1705.00766}.

\bibitemdeclare{phdthesis}{srinivasan2007}
\bibitem{srinivasan2007}
\bibinfo{author}{S.K. \surnamestart Srinivasan\surnameend}
  (\bibinfo{year}{2007}): \emph{\bibinfo{title}{{Efficient Verification of
  Bit-level Pipelined Machines Using Refinement}}}.
\newblock Ph.D. thesis, \bibinfo{school}{Georgia Institute of Technology}.
\newblock \urlprefix\url{http://hdl.handle.net/1853/19815}.

\bibitemdeclare{article}{Vazou:2017}
\bibitem{Vazou:2017}
\bibinfo{author}{Niki \surnamestart Vazou\surnameend}, \bibinfo{author}{Anish
  \surnamestart Tondwalkar\surnameend}, \bibinfo{author}{Vikraman \surnamestart
  Choudhury\surnameend}, \bibinfo{author}{Ryan~G. \surnamestart
  Scott\surnameend}, \bibinfo{author}{Ryan~R. \surnamestart Newton\surnameend},
  \bibinfo{author}{Philip \surnamestart Wadler\surnameend} \&
  \bibinfo{author}{Ranjit \surnamestart Jhala\surnameend}
  (\bibinfo{year}{2017}): \emph{\bibinfo{title}{Refinement Reflection: Complete
  Verification with SMT}}.
\newblock {\sl \bibinfo{journal}{Proc. ACM Program. Lang.}}
  \bibinfo{volume}{2}(\bibinfo{number}{POPL}), pp.
  \bibinfo{pages}{53:1--53:31}, \doi{10.1145/3158141}.

\end{thebibliography}
\end{document}